\documentclass{llncs}
\usepackage{epsfig}
\usepackage{epstopdf}
\usepackage{float}
\usepackage{graphics}
\usepackage{amsmath}
\usepackage{algorithm}
\usepackage{algorithmic}
\usepackage{hhline}

\newcommand{\be}{\begin{eqnarray}}
\newcommand{\ee}{\end{eqnarray}}
\DeclareMathOperator*{\argmax}{arg\,max}
\title{ReLiC: Entity Profiling by using Random Forest and Trustworthiness of a Source - Technical Report}

\author{Shubham Varma, Neyshith Sameer, C. Ravindranath Chowdary } 
 \institute{Department of Computer Science and Engineering, Indian Institute of Technology (BHU) Varanasi, India - 221005 \\{ shvarma@microsoft.com,\,nesameer@microsoft.com,\,rchowdary.cse@iitbhu.ac.in}}
 
\begin{document}

\maketitle

\begin{abstract}

The digital revolution has brought most of the world on the world wide web. The data available on WWW has increased many folds 
in the past decade. Social networks, online clubs and organisations have come into existence. Information is extracted from these venues about a real world entity like a person, organisation, event, etc. However, this information may change over time, and there is a need for the sources to be up-to-date. Therefore, it is desirable to have a model to extract relevant data items from different sources and merge them to build a complete profile of an entity (entity profiling). Further, this model should be able to handle incorrect or obsolete data items. In this paper, we propose a novel method for completing a profile. We have developed a two phase method-1) The first phase (resolution phase) links records to the queries. We have proposed and observed that the use of random forest for entity resolution increases the performance of the system as this has resulted in more records getting linked to the correct entity. Also, we used trustworthiness of a source as a feature to the random forest. 2) The second phase 
selects the appropriate values from records to complete a profile based on our proposed selection criteria. We have used various metrics for measuring the performance of the resolution phase as well as for the overall ReLiC framework. It is established through our results that the use of biased sources has significantly improved the performance of the ReLiC framework. Experimental results show that our proposed system, ReLiC outperforms the state-of-the-art. 


\end{abstract}
\section{Introduction}
\label{INT}
\subsection{Motivation}
Entity profiling is a very challenging problem and has a large number of practical applications. Entity profiling can be used to obtain information about a real-world entity using knowledge bases or by extracting information from unstructured data sources. However, many of the publicly available data sources are not reliable and contain outdated or erroneous data. The challenge is to design a system that works well in the presence of erroneous as well as ambiguous data records. Through entity resolution, relevant data can be linked to the corresponding entity and this data can be used for various information extraction tasks on the entity. Entities can be of different types, for example, people, places, items, etc. Data having different kinds of attributes need to be used for performing entity resolution on different entities. Entity resolution is an important step for entity profiling and to the best of our knowledge, classifiers were not used in the literature for entity resolution. This motivated us to 
study the use of classifiers for entity resolution. Due to which, the proposed method worked effectively on different 
types of data. Entity profiling helps in acquiring accurate information from data extracted from various unreliable sources. But, to the best of our knowledge, assigning a bias to a reliable source for profiling was not considered earlier in the literature and this motivated us to use a biased source. The objective of this work is to develop a novel technique for effectively obtaining the complete profiles of entities using structured data records from different sources.

\subsection{Introduction to Entity Profiling}
Entity Profiling refers to extracting complete information about an entity using information available from different sources that may have incomplete, complete or obsolete data. The entity may be a person, place, object, organization, etc. Each of these entities has their set of attributes, and the values for these attributes are extracted from various online sources. The information provided by these sources may be incomplete, obsolete or even incorrect. Many firms collect data from different sources and create knowledge repositories on real-world entities. Knowledge bases such as Instant Checkmate\footnote{https://www.instantcheckmate.com/} provide complete information about a person by using online profiles. DBPedia \cite{dbpedia1} and YAGO \cite{yago} are other knowledge bases that provide public databases on real-world entities.

\begin{table}[H]
\centering
 \caption {Table of queries} \label{tab:title1} 
\begin{center}
 \begin{tabular}{|c |c | c | c | c | c |} 
 \hline
 query &cluster &Name & Matches & Runs & Highest \\ [0.9ex] 
 \hline\hline
 q1 & c1 & Gavaskar &  & 10122 & \\
 \hline
 q2 & c2 & Amarnath & 69 &  &  \\
 \hline
 q3 & c3 & Y Singh &  &  & 169 \\
 \hline
\end{tabular}
\end{center}
\end{table}

\begin{table}[H]
\centering
 \caption {Table of records} \label{tab:title2} 
\begin{center}
 \begin{tabular}{|c | c | c | c | c | c|} 
 \hline
 record &Name & Matches & Runs & Highest & Source\\ [0.9ex] 
 \hline\hline
 r1 & SM Gavaskar & 125 & 10122 & 236 & s1  \\ 
 \hline
 r2 & Lala Amarnath & 24 &878 & 118 & s1  \\
 \hline
 r3 & Yajurvindra Singh & 4 & 109 &43 & s1\\
 \hline
 r4 & Gavaskar & 125 & 10122 &236 & s2\\
 \hline
 r5 & Mohinder Amarnath & 69 & 4378 &138 & s2 \\  
 \hline
 r6 & Yuvraj Singh & 40 &1900 & 169 & s2\\
 \hline
 r7 & Sunil Gavaskar& 125 & 10122 & 236 & s3\\
 \hline
 r8 & Surinder Amarnath & 10 & 550 & 124 & s3\\
 \hline
 r9 & Yograj Singh & 1 & 10 & 6 & s3\\
 \hline
 r10& Y Singh & 40 & 1900 & 169 & s4\\
 \hline
\end{tabular}
\end{center}
\end{table}

Let us consider Table \ref{tab:title1} which contains three queries. These queries are incomplete profiles and have to be completed using Table \ref{tab:title2}. Table \ref{tab:title2} contains records along with their respective sources. For completing a profile, we must make a cluster of all records that are similar to the query. The clustering of similar records to the query is known as entity resolution.

\begin{table}[H]
 \caption {Table of cluster c1} \label{tab:title3} 
\begin{center}
 \begin{tabular}{|c | c | c | c | c | c|} 
 \hline
 record &Name & Matches & Runs & Highest & Source\\ [0.9ex] 
 \hline\hline
 r1 & SM Gavaskar & 125 & 10122 & 236 & s1  \\ 
 \hline
 r4 & Gavaskar & 125 & 10122 &236 & s2\\
 \hline
 r7 & Sunil Gavaskar& 125 & 10122 & 236 & s3\\
 \hline
\end{tabular}
\end{center}
\end{table}

\begin{table}[H]
 \caption {Table of cluster c2} \label{tab:title4} 
\begin{center}
 \begin{tabular}{|c | c | c | c | c | c|} 
 \hline
 record &Name & Matches & Runs & Highest & Source\\ [0.9ex] 
 \hline\hline
 r2 & Lala Amarnath & 24 &878 & 118 & s1  \\
 \hline
 r5 & Mohinder Amarnath & 69 & 4378 &138 & s2 \\  
 \hline
 r8 & Surinder Amarnath & 10 & 550 & 124 & s3\\
 \hline
\end{tabular}
\end{center}
\end{table}

\begin{table}[H]
 \caption {Table of cluster c3} \label{tab:title5} 
\begin{center}
 \begin{tabular}{|c | c | c | c | c | c|} 
 \hline
 record &Name & Matches & Runs & Highest & Source\\ [0.9ex] 
 \hline\hline
 r3 & Yajurvindra Singh & 4 & 109 &43 & s1\\
 \hline
 r6 & Yuvraj Singh & 40 &1900 & 169 & s2\\
 \hline
 r9 & Yograj Singh & 1 & 10 & 6 & s3\\
 \hline
  r10& Y Singh & 40 & 1900 & 169 & s4\\
 \hline
\end{tabular}
\end{center}
\end{table}

Let us take the first query, i.e, $q1$ having its associated cluster as $c1$. $q1$ refers to the records given in Table \ref{tab:title3}. So, we associated all the records in the Table \ref{tab:title3} to the cluster $c1$. From these records, we complete the profile for $q1$, and the completed profile is given in Table \ref{tab:title6}. Similarly, after applying entity resolution to $q3$, we get Table \ref{tab:title5}. In Table \ref{tab:title5} we see that query can only refer to $r6$ and $r10$ based on the attribute ``Highest". These two records are now used to complete the profile. Table \ref{tab:title6} gives completed profiles. Similar assignment holds for query $q2$.

\begin{table}[H]
 \caption {Table of completed profiles} \label{tab:title6} 
\begin{center}
 \begin{tabular}{| c | c | c | c | c | c|} 
 \hline
 Query &Name & Matches & Runs & Highest\\ [0.9ex] 
 \hline\hline
 q1 & Gavaskar& 125 & 10122 & 236\\
 \hline
 q2 & Amarnath & 69 & 4378 &138\\
 \hline
 q3 & Y Singh & 40 &1900 & 169\\
 \hline
\end{tabular}
\end{center}
\end{table}

Entity resolution is one of the key steps in entity profiling. It involves identifying all the records that refer to a particular entity from knowledge bases published by various sources. It is also referred to as the problem of record linkage, as different records relevant to a given query need to be identified. The similarity between a record and the query is computed to accomplish this task. In the final step, a query is completed by merging the information from the selected set of records.

In this paper, we present a novel technique for completing the profile of an entity by extracting information from different sources. We have used classifiers for linking records to the query. A classifier classifies each record as relevant or non-relevant corresponding to each query. The various input features and labels used for classification are discussed in Section \ref{res_phase}. Section \ref{selclf} describes the technique used for selection of an appropriate classifier model. We have also proposed a novel technique for finding trustworthiness of the sources, to deal with sources that provide erroneous or obsolete data. All the sources are assigned rating with respect to a biased source. The user selects this biased source as the most accurate and relevant source of information. For example, Wikipedia has its information contributed by the wisdom of masses and information on the wiki page needs a valid citation. From time to time, this information is updated as well as maintained. Moreover, different 
knowledge bases like IMDB are entirely dedicated to collect information in their domains. Therefore, it is appropriate to select them as the biased sources for their domains. Further, we have proposed a new metric for finding similarity between two records and between a record and a query. The similarity between two attribute values is dependent on their respective types. We follow different approaches for finding similarities of numerals and strings. This technique is discussed in Section \ref{sim:sec}. There are three major contributions in this work.

\begin{enumerate}
\item We have used classifiers for associating records to a query.
\item We have proposed the use of biased sources for computation of source ratings.
\item We have proposed a novel method for finding similarity between two records and between a record and a query.
\end{enumerate}

The rest of the paper is organized as follows. Section 2 presents the related
work. Terms and definitions are introduced in Section 3. Section 4 outlines the proposed approach. Section 5 describes the experimental results, and we conclude in Section 6.

\section{Related Work}

Entity profiling is a challenging and an active research problem. Li, F. et al. \cite{epvsr} proposed the COMET framework for entity profiling. They have assigned reliability values to every source-attribute pair. They used a two-phase method, involving confidence based matching and adaptive matching. The confidence based matching phase focuses on associating the records to queries and sources. The adaptive matching phase prunes the records linked to multiple queries for reducing the profiling error. A novel technique for profiling entities is proposed in \cite{temporalprofiling}, which tries to model the evolution of an entity over time. In \cite{rluvec}, the focus is on entity profiling by considering uniqueness constraints and false values. They reduce the problem of entity profiling to a k-partite graph clustering problem by performing both global linkage and fusion simultaneously. An ontology based approach for generation of user profiles using YAGO \cite{yago} is proposed in \cite{Calegari2013640}.

The approaches to entity resolution can be classified into two classes - learning based and rule based approaches. A learning-based model for entity resolution using Markov Logic is proposed in \cite{markov}. FEBRL \cite{febrl} (Freely extensible biomedical Record Linkage) uses SVM to learn the appropriate matching combinations. The similarity measures used for this approach is same as that in the rule-based approach proposed in \cite{fellegi1969theory}. MARLIN \cite{addlssm} (Multiply Adaptive Record Linkage with INduction) uses edit distance and cosine similarity measures along with different classifiers for measuring string similarity. A conceptual semantic framework for entity resolution is proposed in \cite{enres}. Zhao, G. et al. \cite{Zhao2016} propose a novel model for linking entities mentioned in texts to semi-structured knowledge bases, by considering relationships between entities which occur together more frequently in text. This technique has been shown to disambiguate entities effectively. 
Cheng, J. et al. \cite{Cheng:2016:LOS:2911451.2914698} have used entity resolution for linking social networking profiles to the organisations the particular user belong to. An analysis for information extraction techniques for linking tweets to entities is has been reported in \cite{Derczynski201532}. Detailed description and analysis of various entity resolution and data linkage approaches are discussed in \cite{book1,book2}.

An entity resolution model that focuses on an incremental approach to building clusters is presented in \cite{irl}. This model merges new records into clusters and modifies the errors based on the new updates. The two approaches proposed by authors are -- connected component approach and iterative approach. These two approaches reduce linkage time without compromising on the quality. Syed, H. et al. \cite{drmrer} discuss refinements in matching rules for the entity resolution using a two-phase model. The first phase computes both the primary attributes of the entity and the baseline matching rules. The second phase involves augmenting these rules to refine and to reduce both false positives and false negatives. It is a rule-based approach and considers the attributes for matches. This approach is very much dependent upon the process of evolving right rules for resolution. Relational clustering for entity resolution is proposed in \cite{cerrd}. Here, both attribute value and the relational information is used 
for associating a record with an entity. Liu, X. et al. \cite{Liu2013264} propose a method for clustering tweets which share a common topic. The Linear Conditional Random Fields model is used for labeling tweets, and then similar tweets are clustered together using these labels.

Xiao, C. et al. \cite{esjndd} proposed a model to join multiple records by the similarity between the corresponding values of a given attribute. The two primary methods incorporated by this approach are - 1) application of similarity function over attribute values and 2) to declare a threshold above which it should be considered as a match. A rule-based approach proposed in \cite{fellegi1969theory} uses the following three similarity measures to compute similarity -- Winkler, Tokenset and Trigram. The similarity threshold is of two types i.e. upper and lower. Anything above the upper threshold means a proper match and below lower threshold means a non-match and between them is a possible match. The idea of quantitative similarity between attribute values is one of the fundamental techniques for entity resolution, which we have incorporated in our proposed method for entity profiling. However, we modified it to suit our model and purpose. A novel method for retrieval of relevant blog posts corresponding to a 
query about an entity by using a facet based information retrieval model is proposed in \cite{Vechtomova201071}.

Benjelloun, O. et al. \cite{swoosh} discusses the pairwise resolution of the entities. The novelty of this approach is the usage of the match and merge properties. The four properties considered here are Idempotence, Associativity, Commutativity and Relativity. The validity of these properties, once confirmed, helps in efficient entity resolution. Bilgic, M. et al. \cite{ddupe} proposed a method to solve entity resolution problem with Markov logic. This method proposes combining of first-order logic and probabilistic graphical models. Weights are associated with first-order formulas and taken as features for Markov networks. The combining of first-order logic and Markov network leads to proper learning and efficient solution to the problem. A theoretical framework for knowledge-based entity resolution using first-order logic is proposed in \cite{Schewe2014101}. The focus is on the analysis of knowledge patterns for optimizing a knowledge model, which is then used for entity resolution. A method for linking 
words or phrases in unstructured texts to entities using part of speech patterns in text is proposed in \cite{Zhang:2015:TEC:2740908.2742766}.

Benny, S. et al. \cite{Benny2016550} proposed a Hadoop framework for entity resolution using the Map and Reduce algorithm on big data. A technique for carrying out entity resolution on heterogeneous distributed probabilistic data is proposed in \cite{Dharavath201593}. They use the expectation maximization algorithm for integrating the data. They have reported significant performance improvements over existing methods for entity resolution in a distributed environment. Ayat, N. et al. \cite{Ayat2014492} proposed a method for entity resolution on probabilistic data. They have proposed algorithms for entity resolution using context-free and context-sensitive similarity functions. Hu, W. et al. \cite{Hu20151} have proposed a scalable technique to address the problem of entity linkage on the semantic web. They have used a bootstrapping method, taking into consideration both the semantic co-referent entities as well as the similarity between property values of the entity for entity resolution.

Wang, J. et al. \cite{crowder} have proposed a hybrid technique for entity resolution using both a human and a system. A system does the initial processing of data, and the users have to choose the correct pair from the most likely pairs identified by the system. Cheng, G. et al. \cite{Cheng2015203} have worked on semi-automatic data integration. They have proposed a technique for selection of features to be used for interactive entity resolution, which makes use of human users to carry out the task of entity resolution. The features are selected in such a way that they convey the largest amount of diverse and characteristic information about an entity. However, this may slow down the entity profiling process, and the performance of the system will be dependent on the knowledge of the people involved. The FEVER framework \cite{ceeraf} is used to analyze different entity resolution approaches. 

\section{Terms and Definitions}
\label{method}

\subsection{Similarity} 
\label{sim:sec}
\subsubsection{Between Record and Query}  
For finding the similarity between a record and a query, the similarity between the corresponding attributes needs to be computed. If the attributes are numeric in nature, then we use percentage difference for assigning the similarity. In the case of strings, we use the distributed representations of words and phrases \cite{word2vec}. A pre-trained model\footnote{GoogleNews-vectors-negative300.bin.gz}, which contains 300 dimension vectors for 3 million words and phrases is used. For this purpose, we used the Word2Vec\footnote{http://code.google.com/archive/p/word2vec} tool for computing similarity. If the words are not found in this model, we use Levenshtein distance for measuring the similarity.
\newline Consider finding the example of similarity between $q1$ in Table \ref{tab:title1} and $r4$ in Table \ref{tab:title2}. The similarity for first attribute values(``\textit{Gavaskar}" and ``\textit{Gavaskar}") is 1.0. As the second attribute is $NULL$ in the query, the similarity is very small, i.e., 0.0001. For the third attribute, the values are identical which makes similarity equal to 1.0. Again the last attribute is $NULL$ in the query. Thus, the similarity value is 0.0001. By summing the similarity of all the attributes, we get 2.0002(1.0 + 0.0001 + 1.0 + 0.0001). Therefore, the similarity between the $q1$ and $r4$ is 2.0002.

\subsubsection{Between Records}
For finding similarity between two records, we have used the same procedure as above. Calculating similarity between record $r6$ and $r10$ in Table \ref{tab:title2} requires calculation of similarity for every attribute value. The first attributes (``$Yuvraj Singh$" and ``$Y Singh$") have a similarity of 0.746. The second, third and fourth attributes have a similarity of 1. By summing the similarity of all the attributes, we get 3.746(0.746 + 1.0 + 1.0 + 1.0). Thus, the similarity between $r6$ and $r10$ is 3.746.

\subsubsection{Between Sources}
The similarity between the sources is used in Section \ref{sec:trust} for calculating the $trustworthiness$ of a source. Let $R_1$ and $R_2$ be two sets of records published by sources $S_1$ and $S_2$ respectively. Then, the similarity between $S_1$ and $S_2$ is given by Equation \ref{eq:src_sim}.
\be
Source\_Similarity(S_1,S_2)=\frac{\sum_{r \in R_1} \max\limits_{r' \in R_2}(similarity(r,r'))}{|R_1|} \label{eq:src_sim}
\ee
For example, let us compute the similarity between sources $s4$ and $s1$, given in Table \ref{tab:title2}. The similarity of $r10$ with $r1$, $r2$ and $r3$ is 0.22, 0.28 and 0.46 respectively. The maximum value is 0.46. Since $s4$ has only one record, the similarity between $s4$ and $s1$ is 0.46. If $s4$ were to have more records, we will repeat this procedure for each record in $s4$, and the similarity between $s4$ and $s1$ will be sum of all the maximum values obtained divided by the number of records in $s4$.

\subsection{Eligibility}
\label{elig}
Eligibility of a record with respect to a cluster is defined as a condition according to which the record can be associated with a given cluster. This association is determined at run time. The eligibility can be decided by some rules or can be determined by using a learning model. Effective identification of conditions can improve the overall accuracy of an entity profiler. Our proposed model identified most of the records related to the given query. The challenge here is to associate correct values to the missing attributes of a query from these records. The overall accuracy of a system is dependent on this step. If wrong records are selected, then in profiling phase, this will impact the selection of values to attributes.
\newline The use of similarity threshold is discussed in \cite{epvsr}. However, there is no standard procedure for selection of the eligibility criteria. The value of this threshold cannot be kept high, as there can be a case where no record will get associated. At the same time this cannot be kept low, as the wrong records may get selected for a cluster. Therefore, the threshold value selection is a non-trivial and a challenging issue in resolution phase. Due to the difficulty in the later approach, we have adopted learning model in this paper. The classifiers are trained to identify the records accurately. Given the training data, they distinguish the genuine records from the non-genuine ones.
\newline $\:$ \\
\subsection{Source rating}
\label{SR}
Source rating is a value associated with a given source, which is a quantitative measure of its trustworthiness. Since trustworthiness is relative, so are the values in the source ratings. A source \textit{$S_1$} (a biased source) can arbitrarily be given a value and other sources will assume source rating values depending on the similarity with \textit{$S_1$}.
\newline We have organized our sources into indexes. Therefore, instead of names we identify the sources by their indexes. After assigning the rating to indexes, we find the source with the maximum rating. The index of this source is termed as the index of source with maximum rating. For example, let $s1$, $s2$ and $s3$ be the sources. Consider the source $s1$ as the biased source, with the value 2.0 assigned to it. The similarity of $s1$, $s2$ and $s3$ with $s1$ is 1.0, 0.4 and 0.5 respectively. The rating of other sources will now be decided according to the source of maximum rating. Therefore, the rating for $s2$ is (2.0 / 1.0 * 0.4)=0.8 and $s3$ is (2.0 / 1.0 * 0.5)=1.0.
\subsection{Attribute Value Set}
\label{AVS}
After linking records to a query in resolution phase, a set $attribute\_value\_set_i$ is defined for each attribute $i$. The set $attribute\_value\_set_i$ will have all the values of attribute $i$ from the selected records. We call these sets as attribute value sets.  Table \ref{tab:title5} gives us records that can be linked to $q_3$. We form attribute value sets for each attribute. The attribute value set $\left\lbrace\textit{4,40,1,40}\right\rbrace$ is for the attribute ``Matches". Likewise, The attribute value set $\left\lbrace\textit{109,1900,10,1900}\right\rbrace$ is for the attribute ``Runs".

\subsection{Source-Similarity Matrix}
\label{SSM}
The source similarity matrix gives the similarity between the sources. If we visualize every source as a vertex and edges be the similarity between the sources, then the source-similarity matrix is an adjacency matrix representation of the graph. The source-similarity matrix will be denoted as \textit{$src\_sim\_mat$} in the algorithms proposed. The source similarity matrix obtained by applying Algorithm \ref{alg:rAP2} on Table \ref{tab:title2} for sources $s1$,$s2$,$s3$ and $s4$ is as follows.
\[
\begin{bmatrix}
-&s1&s2&s3&s4\\
s1&1.0000&0.3141&0.2564&0.1602\\
s2&0.3199&1.0000&0.2868&0.2124\\
s3&0.2803&0.2968&1.0000&0.1802\\
s4&0.2277&0.4671&0.2380&1.0000
\end{bmatrix}
\]

\subsection{Trustworthiness}
\label{sec:trust}
It is an indicator of the correctness and completeness of the records published by a source. The source that is more similar to other sources is considered to be the most trustworthy source as given in Equation \ref{eq1}. $source\_similarity(x,y)$ is the similarity between the sources $x$ and $y$, as discussed in Section \ref{sim:sec}. Let n be the number of sources. We consider the source which is most similar to all other sources as the central or the most trustworthy source $S_{MTS}$, where $MTS$ is the index of the most trustworthy source. The trustworthiness of a source $S_i$ is computed by taking its similarity with the most trustworthy source, as given in Equation \ref{eq2}.
\be
MTS = \argmax\limits_{1 \leq i \leq n}\:(\sum_{j = 1} ^n\,source\_similarity(S_i,S_j))\, \label{eq1} \\
Trustworthiness(S_i) = source\_similarity(S_i,S_{MTS}) \label{eq2}
\ee \\
Let us consider the sources in Table \ref{tab:title2} and the corresponding source similarity matrix given in Section \ref{SSM}. Using Equation \ref{eq1}, we first compute the sum of similarity of each source with all the sources. This value comes out to be 1.73, 1.82, 1.76 and 1.93 for sources $s1$, $s2$, $s3$ and $s4$ respectively. The source $s4$ is selected as the most trustworthy source ($S_{MTS}$). The trustworthiness of a source is its similarity with $s4$. Thus, the trustworthiness of sources $s1$, $s2$, $s3$ and $s4$ are 0.16, 0.21, 0.18 and 1.0 respectively.

\subsection{Similarity-Frequency Product of an Item}\label{SFP}
\label{sim-freq}
Let $R$ be the set of records associated to a query $q$. For an attribute $a$ of $R$, let $A$ be the set of distinct values that the records in $R$ can take on $a$. We define the similarity-frequency product for each value $v \in A$ as follows. Let $S$ denote the similarity between $q_a$ and $v$, where $q_a$ refers to the value of attribute $a$ for a query $q$. Let $F$ denote the number of records in $R$ of attribute $a$ having value $v$. Let $T$ be the sum of similarity of $v$ with all other values in $A$. The similarity-frequency product of $v$ is computed as the product of $S$, $F$ and $T$. It is used in Section \ref{SP} for selecting the correct value for an attribute from a set of possible values.

\section{R\lowercase{e}L\lowercase{i}C: Two-Phase Entity Profiling Framework}
\subsection{Introduction}
There are two phases defined in our model. The first phase is the resolution phase. This phase links the records referring to a query. A record is linked to a query if it is an eligible record as explained in Section \ref{elig}. After resolution, the next phase is selection phase, where we compute the values of attributes from these records.
\subsubsection{Resolution Phase}
\label{res_phase}
The resolution phase associates the records with a given query. We employ a learning model for classifying the records that should be linked to a query. The performance of this phase is crucial to the overall performance of the system. A suitable classifier is selected for this phase (details in Section \ref{selclf}). Since the primary purpose of the classifier is to classify the records as relevant or non-relevant corresponding to a query, the input features should capture the information related to the record and the query. Corresponding to each attribute, we consider the similarity between the query and the value in the record, as described in Section \ref{sim:sec}, as an input feature for our model. A record may not have the complete information about an entity, i.e., some attribute values may be missing. To capture this in our model, for each attribute, we have an input feature which can take values 0 or 1, where 0 corresponds to a missing value. Further, we also consider the trustworthiness of the 
source, as described in Section \ref{sec:trust}, as an input feature for the classifier.
Let us take the following query $q$ and record $r$, where the attributes correspond to restaurant name, phone number, website and city respectively.
\begin{center}
\textit{r=$<$Pizza Point, 909476941, NULL, Varanasi$>$\\
q=$<$Pizza Corner, 785561264, NULL, NULL$>$}
\end{center}
Here, the similarity between the corresponding attributes is 0.36\footnote{Note that this value is computed using Word2Vec}, 0.14, 0.0 and 0.0 respectively. Note that similarity for the last two attributes is 0.0 due to the missing values in record or query. These similarity values are an input feature for the classifier. Since the record has the third attribute as NULL, the feature that captures the missing information in the record, corresponding to each attribute, will take values 1, 1, 0 and 1 respectively. Let the trustworthiness of the source that published $r$ be 0.62. In this case, the input feature to the classifier will be (0.36, 0.14, 0.0, 0.0, 1, 1, 0, 1, 0.62).

The labels are obtained for each record query pair using the ground truth data. With these input features and labels, we train the classifier using 80\% of the queries and the remaining queries are used for testing the system.

\subsubsection{Selection Phase}
\label{SP}
The records linked to a query are selected and processed to complete the profile. The values used in a profile are selected from the linked records. We compute different attribute value sets as described in Section \ref{AVS}. Consider an example of $q1$ in Table \ref{tab:title1}. After entity resolution for $q1$, we get Table \ref{tab:title3}. Table \ref{tab:title3} gives us records that can be linked to $q_1$. We form attribute value sets for each attribute. The attribute value set \textit{SM Gavaskar, Gavaskar, Sunil Gavaskar} is for the attribute ``Name".
\newline In this phase, we compute similarity-frequency product as discussed in Section \ref{SFP}. For a given attribute value, it is given as the product of similarity with the query value for that attribute, it's frequency and sum of similarity with other attribute values in that attribute. For example, let us take ``Name`` attribute value set for query $q1$. Let us compute similarity-frequency product for attribute value $Gavaskar$. Let the similarity of $Gavaskar$ with \textit{SM Gavaskar} be 0.7 and \textit{Sunil Gavaskar} be 0.6. The sum of similarity is 1.30. Since the query value is also \textit{Gavaskar}, therefore similarity is 1. The frequency of $Gavaskar$ in attribute value set of ``Name" is 1. Therefore, its similarity-frequency product is (0.6+0.7)*1*1, i.e., 1.30. Now, two different values are obtained. First is the similarity to the source with maximum source rating and second is similarity-frequency product. The attribute value whose product of two values is maximum is chosen to be assigned 
to 
the attribute. If \textit{Sunil Gavaskar } has the product of these two terms as maximum then it will be assigned as the value for attribute ``Name". The other attribute value sets have only one element. The attribute value set for ``matches" has $125$, ``Runs" has $10122$ and ``Highest" has $236$. Therefore, other attributes in profile have to assume the values in their attribute value set respectively. The final profile for $q1$ is shown in Table \ref{tab:title6}.

\subsection{Proposed Algorithms}
Algorithm \ref{alg:rAP1} gives the procedure to compute the similarity between two records. A variable $sim\_value$ is used to store the result. We have two values for each attribute, one by record $a$ and the other by record $b$. Both these values are given as input to $similar(x,y)$ function. We compute the sum of values returned by similarity function for each attribute. The similar function is polymorphic in nature. The choice of which version to use is dependent on the type of the attribute(string or real number).
\newline Let us compute the similarity between two records $r1$ and $r4$ in Table \ref{tab:title2}. The records have four attributes i.e. ``Name", ``Matches", ``Runs", ``Highest". The similarity between the attribute values \textit{SM Gavaskar} and \textit{Gavaskar} be 0.70. Rest of the attribute values(for attributes ``Matches", ``Runs" are ``Highest") are identical. Therefore, the similarity between them will be 1.0. Thus, the similarity between the records will be 0.70 + 1.0 + 1.0 + 1.0 = 3.7.
\begin{algorithm}
\textbf{Input:} Records $a$ and $b$ \\
\textbf{Output:} Similarity value between $a$ and $b$ \\
\begin{algorithmic}[1]
	\STATE $sim\_value \gets 0$
    \FOR {each attribute $i$}
		\STATE $sim\_value$ += $similar(a_i,b_i)$\COMMENT{Details in Section \ref{sim:sec}}
    \ENDFOR 
    \RETURN $sim\_value$
\end{algorithmic}
    \caption{Similarity Function}
	\label{alg:rAP1}
\end{algorithm}
\newline
Algorithm \ref{alg:rAP2} is to compute the source similarity matrix. Our task is to form a graph, where sources are represented by vertex and similarity between them is represented by an edge between them. The similarity between the two sources is given by the sum of similarities between their records. The similarity function in Algorithm 2 takes two records as input. First record is from $s_1$ and second from $s_2$. We generate a matrix, with rows and columns representing the sources, and the values representing similarity between the two sources. Each element in the matrix is divided by the product of the number of records in the two sources .
\begin{algorithm}
\textbf{Input:} Set of sources $S$ \\
\textbf{Output:} Source similarity matrix $src\_sim\_mat$ \\

    \begin{algorithmic}[1]
		\FOR {every source $s_1$}
			\FOR {every source $s_2$}
				\STATE \text{$src\_sim\_mat_{s_1,s_2} \gets 0$}
				\FOR {every record $r_1 \in s_1$}
				\STATE \text{$maxsim \gets 0$}
					\FOR {every record $r_2 \in s_2$}
				\STATE \text{$maxsim  \gets $ max$(maxsim,similarity(r_1,r_2))$}
				\ENDFOR
				\STATE \text{$src\_sim\_mat_{s_1,s_2} += maxsim$}
			\ENDFOR
		\ENDFOR
		\ENDFOR
		\FOR {every source $s_1$}
			\FOR {every source $s_2$}
				\STATE $src\_sim\_mat_{s_1,s_2} \gets \frac{src\_sim\_mat_{s_1,s_2}}{|s_1|}$
			\ENDFOR
		\ENDFOR
    \end{algorithmic}
    \caption{Source Similarity Matrix}
    \label{alg:rAP2}
\end{algorithm}
\newline
Algorithm \ref{alg:rAP3} is for entity profiling. Let us consider the set of records in Table \ref{tab:title2} and query $q1$ in Table \ref{tab:title1}. In lines 1-5, we associate records to $q_1$ using the pre-trained classifier model as discussed in Section \ref{res_phase}. After this, we get all the records linked to $q1$ as given in Table \ref{tab:title3}. Next, we assign rating to all the sources as described in Section \ref{SR}. Let us take three sources -- $s1$, $s2$ and $s3$ having indexes 1, 2 and 3 respectively. Let the ratings assigned to them be 1.0, 2.1 and 0.8 respectively. The source index with maximum rating is assigned to the $index\_of\_maximum$. The $index\_of\_maximum$ in our case is 2, corresponding to $s2$. We generate attribute value set for each attribute, containing all distinct values that the attribute assumes. The attribute value set for the attribute ``Name" is \textit{SM Gavaskar, Gavaskar, Sunil Gavaskar}, for ``Matches" is $\left\lbrace\textit{125}\right\rbrace$, for ``Runs" 
is $\left\lbrace\textit{10122}\right\rbrace$ and for ``Highest" is $\left\lbrace\textit{236}\right\rbrace$. In lines 20-24, we calculate the sum of similarity of an attribute value with other values of that attribute.
This sum, for a given attribute value $v$ and attribute $i$, is denoted by $sim\_attribute\_val_{v,i}$.
We assume that majority of the values in attribute value set have high similarity with each other, and the correct value resides among them. To find this value, we take the sum of similarity of an attribute value with all other values. This way, the majority of the values which are similar to each other get high $sim\_attribute\_val_{v,i}$. Further, the attribute value which is central, i.e., most similar to all other values gets the highest $sim\_attribute\_val_{v,i}$. We use $sim\_attribute\_val$ in the calculation of the similarity-frequency product.
Further, for every value in attribute value set for attribute $i$, we calculate its similarity-frequency product as described in Sections \ref{sim-freq} and \ref{SP}. Finally, the attribute value whose product of the similarity to \textit{$index\_of\_maximum$} and the similarity-frequency product is maximum is chosen for assignment.

\section{Experimental Setup and Results}
In this section, we present the results of the extensive experiments done to validate the proposed methodology.

\subsection{Datasets}
For our experiments, use real world datasets on restaurant, football and cricket. All these datasets have records extracted from multiple sources.

\subsubsection{Restaurant Dataset}
The restaurant dataset in \cite{epvsr} contains information about restaurants with the zip code 78701. It consists of 581 records on 222 restaurants, extracted from 6 websites. Each record has values corresponding to name, address, phone, website and id. The ground truth is obtained from extracting information about these restaurants from Yellow Pages\footnote{www.yellowpages.com}. Each query comprises of two missing attributes and three filled attributes.

\subsubsection{Football Dataset}
The football dataset in \cite{Blanco2012} contains information about football players. It consists of 7492 records extracted from 20 websites. Each record has values corresponding to name, birth date, height, weight, playing position and birth place. The ground truth were extracted from the official sites of the football clubs which these players represent. In \cite{epvsr}, a series of datasets are generated using the original football dataset by introducing errors and ambiguities in the data. The errors are generated by randomly generating an erroneous value and ambiguities are generated by abbreviating names, removing first name or removing the last name. The percentage error is varied from 0.1-0.5, whereas the percentage ambiguity is varied from 0.4-0.9. Each query comprises of three missing attributes and four filled attributes.

\begin{algorithm}[H]
\textbf{Input:} Query $q$, set of records $R$, set of sources $S$ \\
\textbf{Output:} Completed profile $p$ \\
    \begin{algorithmic}[1]
        \FOR{every $r \in R$  }
        	\STATE $s \gets source(r)$ 
            \IF{classifier.predict($r$,$q$,$s$) == 1}
			\STATE            
            \COMMENT{Classifier predicts $r$ as genuine with respect to $q$}
            	\STATE \text{associate $r$ with $q$}
            \ENDIF
        \ENDFOR
        \STATE $src\_rating \gets $ list of ratings of all sources
        \STATE $index\_of\_maximum \gets {max\_index(src\_rating)}$
        \FOR{every attribute $i$}
            \STATE $attribute\_value\_set_i \gets {\phi}$
            \FOR{every record $r$ associated to $q$}
                \STATE $attribute\_value\_set_i \gets attribute\_value\_set_i \cup r_i$
            \ENDFOR
        \ENDFOR
        \FOR {every attribute $i$}
			\FOR {every value $\in attribute\_value\_set_i$}
				\STATE \text{Initialize dictionary} $sim\_attribute\_val_{value} \gets 0$
			\ENDFOR
			\FOR {every $value_1 \in attribute\_value\_set_i$}
				\FOR {every $value_2 \in attribute\_value\_set_i$ and $value_1 \not= value_2$}
					\STATE \textit{$sim\_attribute\_val_{value_1}$ += similar($value_1$, $value_2$)}
				\ENDFOR
			\ENDFOR
			\FOR {every value $v$ in \textit{$attribute\_value\_set_i$}}
				\STATE $V1_{v,i} \gets src\_sim\_mat_{index\_of\_maximum,s}$ \COMMENT{$s$ is the source from which the $v$ comes}
				\STATE $V2_{v,i} \gets similarity\_frequency\_product_{v,i}$ \COMMENT{Explained in Section \ref{SFP}}
				\STATE ${V3_{v,i} \gets V1_{v,i}*V2_{v,i}}$
			\ENDFOR
			\STATE{\textit{The value in $attribute\_value\_set_i$ having the maximum $V3_{v,i}$ is assigned to $p_i$}}
		\ENDFOR
    \end{algorithmic}
    \caption{Entity profiling}
    \label{alg:rAP3}
\end{algorithm}

\begin{figure*}
\centering
        \includegraphics[width=\columnwidth, height=3in]{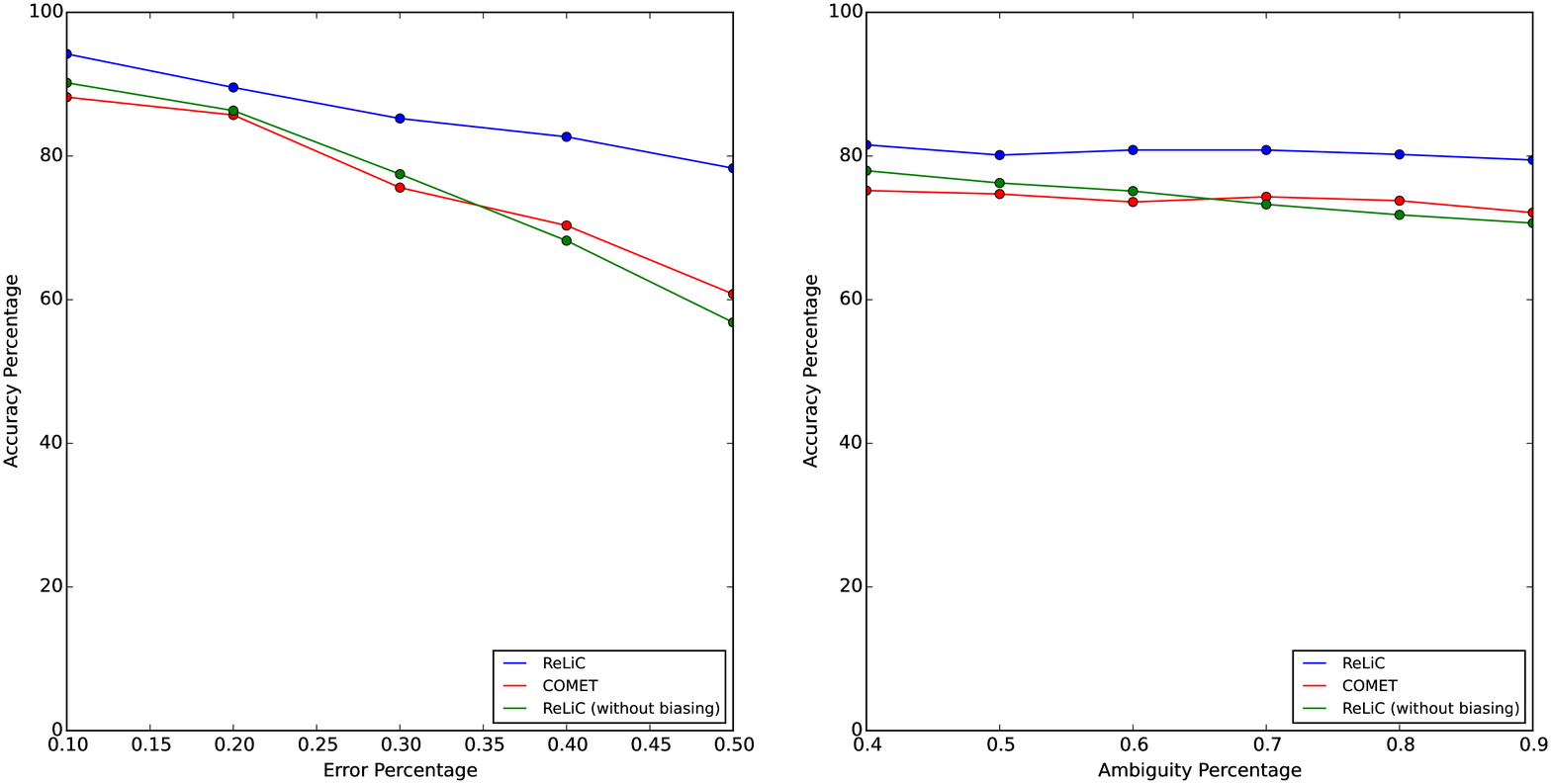}
\caption{Accuracy for Football Dataset}
\label{football1}
\end{figure*}

$\:$
\subsubsection{Cricket Dataset}
We prepared cricket dataset by extracting statistical information on cricketers in Test Cricket. The data comprises of 330 records extracted from 3 websites. Each record has values corresponding to name,the number of matches, net runs, highest score, batting average, number of hundreds, number of fifties and country. The ground truth is obtained from the Wikipedia pages of each of these players. Each query comprises of four missing attributes and four filled attributes.

\subsection{Selection of Classifier Model}
\label{selclf}
As discussed in Section \ref{res_phase}, an appropriate classifier model needs to be selected for assigning records to the queries. For this purpose, we randomly select 80\% of the records as training data and remaining as testing data. We calculate the $F1$ score, cross correlation error, ROC AUC score \cite{rocauc} and the Matthews Correlation Coefficient (MCC) \cite{Matthews1975442} for different classifiers on the testing data. The $F1$ score is the harmonic mean of the precision and recall. The cross-validation error was computed using 10 fold strategy; wherein the data is divided into 10 subsets of equal length. During each iteration, one of the subsets is considered as a test set and the rest are used for training. ROC AUC score refers to the area under the receiver operating characteristic (ROC) curve. The ROC curve is a plot between the recall and the fall-out of the classifier. It is a standard technique used for the comparison of classifier models \cite{rocauccomp}. The Matthews Correlation Score (
MCC) 
is used for measuring the quality of binary classifiers. $TP$, $FP$, $TN$ and $FN$ denotes true positives, false positives, true negatives and false negatives respectively. $MCC$ is computed using Equation \ref{MCC}.

\begin{equation}
\label{MCC}
MCC=\frac{TP * TN - FP * FN}{\sqrt{(TP+FP)(TP+FN)(TN+FP)(TN+FN)}}
\end{equation}

The results of various classifier models is given in Table \ref{table1}. We used scikit-learn \cite{scikit} python library for implementing these classifiers. Note that, we used SVM with linear kernel, and random forest with 10 trees. From the Table \ref{table1}, we can conclude that the Random Forest Classifier clearly outperforms all the other classifiers. Therefore, we have chosen this classifier for obtaining the subsequent results.

\begin{table}[htbp]
\caption{Performance of Classifier Models}\label{table1}
\begin{center}
\begin{tabular}{|c|c|c|c|c|}\hline
{\bfseries Classifier}&{\bfseries F1}&{\bfseries Cross}&{\bfseries ROC}&{}\\
{\bfseries Model}&{\bfseries Score}&{\bfseries Validation}&{\bfseries AUC}&{\bfseries MCC}\\
{}&{}&{\bfseries Error (\%)}&{\bfseries Score}&{}\\
\hline
Naive Bayes&0.8412&6.23&0.8494&0.7295\\
\hline
K Nearest Neighbour&0.8724&4.24&0.9177&0.8362\\
\hline
SVM&0.9088&4.33&0.9315&0.8344\\
\hline
Decision Tree&0.8973&4.15&0.9035&0.8011\\
\hline
Random Forest&0.9567&2.72&0.9511&0.9124\\
\hline
\end{tabular}
\end{center}\end{table}

$\:$
\subsection{Results}

\begin{figure*}
\centering
        \includegraphics[width=\columnwidth, height=3in]{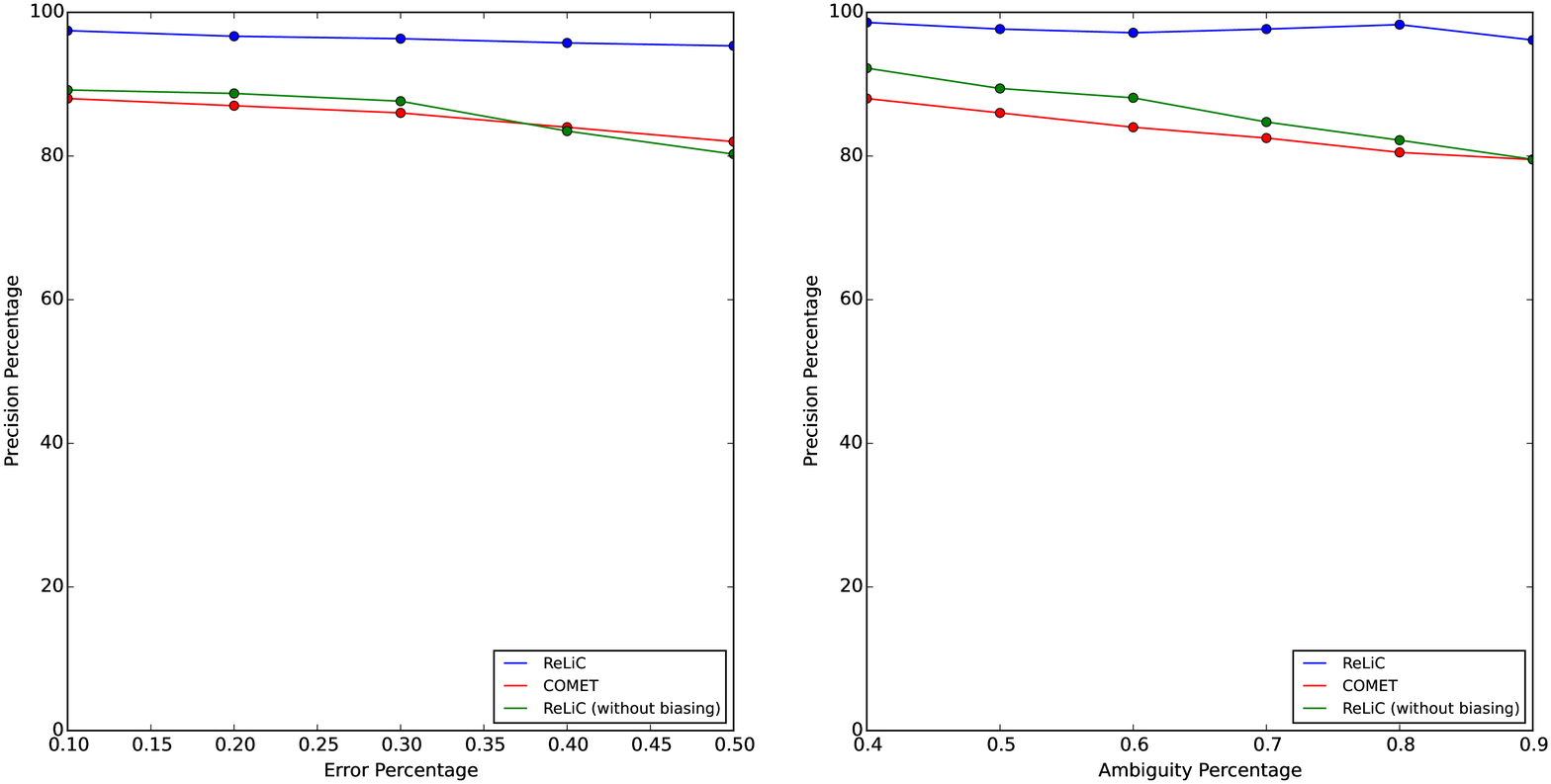}
\caption{Precision for Football Dataset}
\label{football2}
\end{figure*}

We compare ReLiC with the COMET framework \cite{epvsr}, which is the state-of-the-art entity profiling technique. We evaluate our model for entity resolution and also for overall profile completion. Results are evaluated using three metrics - accuracy, precision and recall.

Precision and recall are used for evaluating the entity resolution phase. They are calculated on the records associated with each query using Equations \ref{pre} and \ref{rec}. Let $R_i$ be the set of records associated with query $q_i$ and $R'_i$ be the set of records that get associated to query $q_i$ during the resolution phase of the algorithm.
\be
Precision=\frac{|R_i \cap R'_i|}{|R'_i|}  \label{pre} \\
Recall=\frac{|R_i \cap R'_i|}{|R_i|}  \label{rec}
\ee
The accuracy metric is used for evaluating the entire entity profiling approach. It refers to the amount of similarity between the query filled by the entity profiling approach and the ground truth values corresponding to that query, as seen in Equations \ref{acc} and \ref{accsim}. Here, similarity between the query and ground truth values is calculated by taking the average of the similarity between the corresponding attributes. If the attributes are numerals, then we use the ratio of difference, otherwise, we use Levenshtein distance \cite{levenshtein1966binary} for computing the similarity between two attributes. 
\be
Accuracy =\frac{1}{N} \sum_{i=1}^{N} \frac{1}{|A|} \sum_{a \in A} sim(q_a,t_a) \label{acc}
\ee
Here, $N$ denotes the number of filled queries, $q$ denotes the filled query, $t$ denotes the corresponding truth values and $A$ denotes the set of all attributes.

\begin{equation}
\label{accsim}
sim(v_1,v_2) = \begin{cases}
\frac{|v_1-v_2|}{max(v_1,v_2)} &\text{if $v_1,v_2 \in \rm I\!R$}\\
Levenshtein(v_1,v_2) &\text{otherwise}
\end{cases}
\end{equation}

Here, $Levenshtein(a,b)$ refers to the edit distance between $a$ and $b$.

\begin{table}[htbp]
\caption{Results for Restaurant dataset}\label{rest}
\begin{center}
\begin{tabular}{|c|c|c|c|}\hline
{\bfseries Methodology}&{\bfseries Accuracy}&{\bfseries Precision}&{\bfseries Recall}\\
\hline
\bfseries ReLiC&\bfseries 93.7913&\bfseries 97.1965&\bfseries 96.2475\\
\hline
COMET&86.2474&95.2671&95.1111\\
\hline
ReLiC (without biasing)&82.6437&93.6403&92.4048\\
\hline
\end{tabular}
\end{center}\end{table}

\begin{table}[htbp]
\caption{Results for Cricket dataset}\label{crick}
\begin{center}
\begin{tabular}{|c|c|c|c|}\hline
{\bfseries Methodology}&{\bfseries Accuracy}&{\bfseries Precision}&{\bfseries Recall}\\
\hline
\bfseries ReLiC&\bfseries 87.2733&\bfseries 92.9083&\bfseries 92.0222\\
\hline
COMET&80.9424&88.6667&86.2471\\
\hline
ReLiC (without biasing)&82.2431&88.3534&87.7356\\
\hline
\end{tabular}
\end{center}\end{table}

We have also calculated the accuracy, precision and recall values for ReLiC framework, without the use of biased sources. In this case, we have kept the rating of all the sources to be equal. It is evident that there is a significant impact on the performance of the system with biasing. From this, we can infer that biasing boosts the performance of a profiling system.

\begin{figure*}
\centering
        \includegraphics[width=\columnwidth, height=3in]{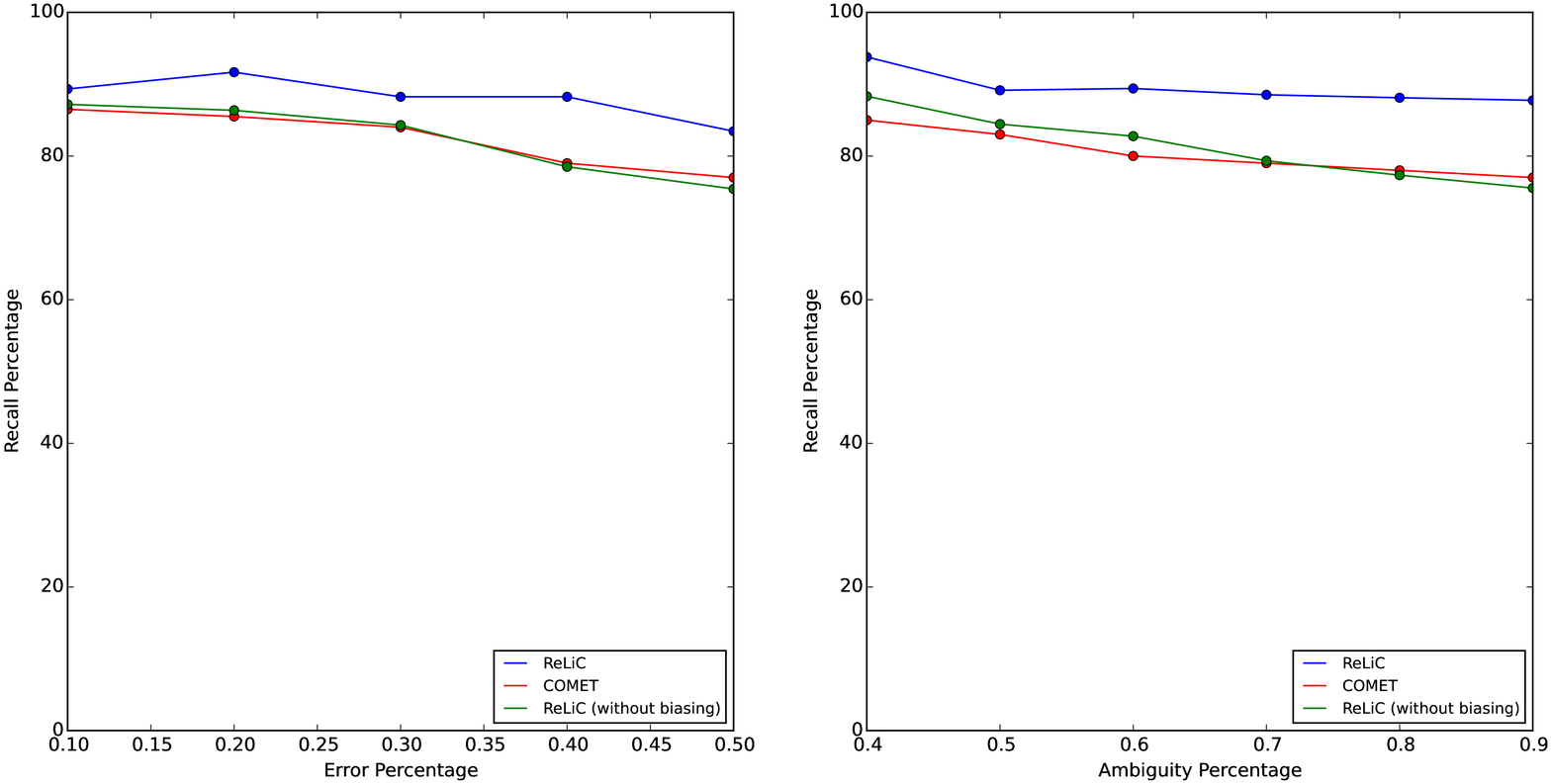}
\caption{Recall for Football Dataset}
\label{football3}
\end{figure*}

For the ReLiC framework, we use 70\% of the queries for training the model, and the remaining queries are used for testing. We calculate the accuracy, precision and recall to test the performance of all the models. Tables \ref{rest} and \ref{crick} show the results for the restaurant and cricket datasets respectively. Figures \ref{football1}, \ref{football2} and \ref{football3} show the performance of all the systems on football dataset. From these results, we can conclude that our proposed system outperforms the current state-of-the-art entity profiling framework.

\begin{table}[htbp]
\caption{Result for 2-sided paired student's t-test on ReLiC and COMET for Restaurant dataset}\label{tab:statrest}
\begin{center}
\begin{tabular}{|c|c|c|c|}\hline
{\bfseries Metric}&{\bfseries T Value}&{\bfseries P Value}&{\bfseries Sample Effect Size}\\
\hline
Precision&2.9600&0.00473&0.9239\\
\hline
Recall&2.8189&0.00693&1.0911\\
\hline
Accuracy&3.5213&0.00094&1.3158\\
\hline
\end{tabular}
\end{center}\end{table}

\begin{table}[htbp]
\caption{Result for 2-sided paired student's t-test on ReLiC and COMET for Football dataset}\label{tab:statfoot}
\begin{center}
\begin{tabular}{|c|c|c|c|}\hline
{\bfseries Metric}&{\bfseries T Value}&{\bfseries P Value}&{\bfseries Sample Effect Size}\\
\hline
Precision&2.5102&0.01542&0.9420\\
\hline
Recall&3.3164&0.00172&1.2563\\
\hline
Accuracy&2.3313&0.02390&1.3589\\
\hline
\end{tabular}
\end{center}\end{table}

\begin{table}[htbp]
\caption{Result for 2-sided paired student's t-test on ReLiC and COMET for Cricket dataset}\label{tab:statcric}
\begin{center}
\begin{tabular}{|c|c|c|c|}\hline
{\bfseries Metric}&{\bfseries T Value}&{\bfseries P Value}&{\bfseries Sample Effect Size}\\
\hline
Precision&3.8382&0.00036&1.0762\\
\hline
Recall&3.0846&0.00334&1.2757\\
\hline
Accuracy&2.5009&0.01578&1.2444\\
\hline
\end{tabular}
\end{center}\end{table}

For validating the statistical significance of our experiments, we have conducted a 2-sided paired student's t-test as described in \cite{Sakai:2014:SRI:2641383.2641385}. For this purpose, we have compared ReLiC with COMET, using all the 3 datasets (cricket, football, restaurant) on each of the performance metrics -- precision, recall and accuracy. We calculate the T value, P value and the sample effect size. The value of $\alpha$ is chosen as 0.05. The values are reported in Tables \ref{tab:statrest}, \ref{tab:statfoot} and \ref{tab:statcric}. From these tables, it is clear that the obtained P values are less than $\alpha$. Further, T values and sample effect sizes are considerable. This shows that ReLiC significantly outperforms COMET.

\subsection{Discussion}
Various knowledge bases are entirely devoted to the collection of data related to their fields. Owners, organisations or followers regularly update these sources. Thus, the information content in these sources is generally accurate. To give more importance to the data published by these sources, we have used the concept of a biased source for entity profiling. The concept of biased source allows such sources to have higher ratings as compared to others. The results demonstrate that the use of biased sources is effective for entity profiling. As seen in Table \ref{rest}, the use of biased source in ReLiC has shown an increase of around 11\% in accuracy and around 4\% in precision and recall.

COMET is the state-of-the-art framework for entity profiling. It uses a clustering-based technique for associating records to queries. Further, it uses pruning for removal of irrelevant records from the queries. On the other hand, ReLiC uses the random forest for associating records to queries. Table \ref{table1} shows that random forest classifier has a high F1 score of 0.9567. Since random forest links most of the relevant records to the queries, pruning is not required. The next step involves selection of value from the set of attribute values obtained. COMET uses source reliability for this purpose, whereas ReLiC uses the similarity-frequency product for selecting the correct attribute value.

\section{Conclusions}
In this paper, we address the problem of entity profiling using a two-phase framework (ReLiC). In the first phase, we assign ratings to the sources by computing their similarity with respect to a biased source. We have used Word2Vec for computing similarity between strings. We associate records to a query using random forest classifier. Our results demonstrate that the usage of a biased source improved the overall performance of ReLiC. We used three similarity measures - query-record, record-record and source-source to address the problem of entity profiling. We compared our system with the state-of-the-art COMET framework and our results clearly demonstrate that ReLiC outperforms COMET. ReLiC framework can be further improved by considering temporal information associated to the records. This can allow the user to get the complete profile of an entity at different points of time. Further, ReLiC and COMET assume that the details given in the user query are accurate. Therefore, considering errors in user 
query may improve the performance of ReLiC.

\bibliographystyle{plain}

\begin{thebibliography}{41}
\bibitem{dbpedia1}
S{\"o}ren Auer, Christian Bizer, Georgi Kobilarov, Jens Lehmann, Richard
  Cyganiak, and Zachary Ives.
\newblock Dbpedia: A nucleus for a web of open data.
\newblock In {\em The Semantic Web: 6th International Semantic Web Conference,
  2nd Asian Semantic Web Conference, ISWC 2007 + ASWC 2007, Busan, Korea,
  November 11-15, 2007. Proceedings}, pages 722--735. Springer Berlin
  Heidelberg, 2007.

\bibitem{Ayat2014492}
Naser Ayat, Reza Akbarinia, Hamideh Afsarmanesh, and Patrick Valduriez.
\newblock Entity resolution for probabilistic data.
\newblock {\em Information Sciences}, 277:492 -- 511, 2014.

\bibitem{swoosh}
Omar Benjelloun, Hector Garcia-Molina, David Menestrina, Qi~Su, Steven~Euijong
  Whang, and Jennifer Widom.
\newblock Swoosh: A generic approach to entity resolution.
\newblock {\em The VLDB Journal}, 18(1):255--276, 2009.

\bibitem{Benny2016550}
S.~Prabhakar Benny, S.~Vasavi, and P.~Anupriya.
\newblock Hadoop framework for entity resolution within high velocity streams.
\newblock {\em Procedia Computer Science}, 85:550 -- 557, 2016.
\newblock International Conference on Computational Modelling and Security (CMS
  2016).

\bibitem{cerrd}
Indrajit Bhattacharya and Lise Getoor.
\newblock Collective entity resolution in relational data.
\newblock {\em ACM Trans. Knowl. Discov. Data}, 1(1):1--36, 2007.

\bibitem{addlssm}
Mikhail Bilenko and Raymond~J. Mooney.
\newblock Adaptive duplicate detection using learnable string similarity
  measures.
\newblock In {\em Proceedings of the Ninth ACM SIGKDD International Conference
  on Knowledge Discovery and Data Mining}, pages 39--48. ACM, 2003.

\bibitem{ddupe}
M.~Bilgic, L.~Licamele, L.~Getoor, and B.~Shneiderman.
\newblock D-dupe: An interactive tool for entity resolution in social networks.
\newblock In {\em Visual Analytics Science And Technology, 2006 IEEE Symposium
  On}, pages 43--50, 2006.

\bibitem{Blanco2012}
Lorenzo Blanco, Valter Crescenzi, Paolo Merialdo, and Paolo Papotti.
\newblock {\em Web Data Reconciliation: Models and Experiences}, pages 1--15.
\newblock Springer Berlin Heidelberg, 2012.

\bibitem{rocauc}
Andrew~P. Bradley.
\newblock The use of the area under the \{ROC\} curve in the evaluation of
  machine learning algorithms.
\newblock {\em Pattern Recognition}, 30(7):1145 -- 1159, 1997.

\bibitem{Calegari2013640}
Silvia Calegari and Gabriella Pasi.
\newblock Personal ontologies: Generation of user profiles based on the
  \{YAGO\} ontology.
\newblock {\em Information Processing \& Management}, 49(3):640 -- 658, 2013.
\newblock Personalization and Recommendation in Information Access.

\bibitem{Cheng2015203}
Gong Cheng, Danyun Xu, and Yuzhong Qu.
\newblock C3d+p: A summarization method for interactive entity resolution.
\newblock {\em Web Semantics: Science, Services and Agents on the World Wide
  Web}, 35, Part 4:203 -- 213, 2015.

\bibitem{Cheng:2016:LOS:2911451.2914698}
Jerome Cheng, Kazunari Sugiyama, and Min-Yen Kan.
\newblock Linking organizational social network profiles.
\newblock In {\em Proceedings of the 39th International ACM SIGIR Conference on
  Research and Development in Information Retrieval}, SIGIR '16, pages
  901--904, New York, NY, USA, 2016. ACM.

\bibitem{book1}
P.~Christen.
\newblock {\em Data Matching: Concepts and Techniques for Record Linkage,
  Entity Resolution, and Duplicate Detection}.
\newblock Springer Berlin Heidelberg, 2012.

\bibitem{febrl}
Peter Christen.
\newblock Febrl: A freely available record linkage system with a graphical user
  interface.
\newblock In {\em Proceedings of the Second Australasian Workshop on Health
  Data and Knowledge Management - Volume 80}, pages 17--25. Australian Computer
  Society, Inc., 2008.

\bibitem{Derczynski201532}
Leon Derczynski, Diana Maynard, Giuseppe Rizzo, Marieke van Erp, Genevieve
  Gorrell, Raphaël Troncy, Johann Petrak, and Kalina Bontcheva.
\newblock Analysis of named entity recognition and linking for tweets.
\newblock {\em Information Processing \& Management}, 51(2):32 -- 49, 2015.

\bibitem{Dharavath201593}
Ramesh Dharavath and Chiranjeev Kumar.
\newblock Entity resolution based \{EM\} for integrating heterogeneous
  distributed probabilistic data.
\newblock {\em Journal of Systems and Software}, 107:93 -- 109, 2015.

\bibitem{fellegi1969theory}
Ivan~P Fellegi and Alan~B Sunter.
\newblock A theory for record linkage.
\newblock {\em Journal of the American Statistical Association},
  64(328):1183--1210, 1969.

\bibitem{irl}
Anja Gruenheid, Xin~Luna Dong, and Divesh Srivastava.
\newblock Incremental record linkage.
\newblock {\em Proc. VLDB Endow.}, 7(9):697--708, 2014.

\bibitem{rluvec}
Songtao Guo, Xin~Luna Dong, Divesh Srivastava, and Remi Zajac.
\newblock Record linkage with uniqueness constraints and erroneous values.
\newblock {\em Proc. VLDB Endow.}, 3(1-2):417--428, 2010.

\bibitem{rocauccomp}
J~A Hanley and B~J McNeil.
\newblock A method of comparing the areas under receiver operating
  characteristic curves derived from the same cases.
\newblock {\em Radiology}, 148(3):839--843, 1983.

\bibitem{Hu20151}
Wei Hu and Cunxin Jia.
\newblock A bootstrapping approach to entity linkage on the semantic web.
\newblock {\em Web Semantics: Science, Services and Agents on the World Wide
  Web}, 34:1 -- 12, 2015.

\bibitem{ceeraf}
Hanna K\"{o}pcke, Andreas Thor, and Erhard Rahm.
\newblock Comparative evaluation of entity resolution approaches with fever.
\newblock {\em Proc. VLDB Endow.}, 2(2):1574--1577, 2009.

\bibitem{levenshtein1966binary}
Vladimir~I Levenshtein.
\newblock Binary codes capable of correcting deletions, insertions, and
  reversals.
\newblock {\em Soviet physics doklady}, 10(8):707--710, 1966.

\bibitem{epvsr}
Furong Li, Mong~Li Lee, and Wynne Hsu.
\newblock Entity profiling with varying source reliabilities.
\newblock In {\em Proceedings of the 20th ACM SIGKDD International Conference
  on Knowledge Discovery and Data Mining}, pages 1146--1155. ACM, 2014.

\bibitem{temporalprofiling}
Furong Li, Mong~Li Lee, Wynne Hsu, and Wang-Chiew Tan.
\newblock Linking temporal records for profiling entities.
\newblock In {\em Proceedings of the 2015 ACM SIGMOD International Conference
  on Management of Data}, pages 593--605. ACM, 2015.

\bibitem{Liu2013264}
Xiaohua Liu and Ming Zhou.
\newblock Two-stage \{NER\} for tweets with clustering.
\newblock {\em Information Processing \& Management}, 49(1):264 -- 273, 2013.

\bibitem{enres}
Bradley Malin and Latanya Sweeney.
\newblock Enres: a semantic framework for entity resolution modelling.
\newblock In {\em Institute for Software Research International Technical
  Report}. Carnegie Mellon University, 2005.

\bibitem{Matthews1975442}
B.W. Matthews.
\newblock Comparison of the predicted and observed secondary structure of
  \{T4\} phage lysozyme.
\newblock {\em Biochimica et Biophysica Acta (BBA) - Protein Structure},
  405(2):442 -- 451, 1975.

\bibitem{word2vec}
Tomas Mikolov, Ilya Sutskever, Kai Chen, Greg~S Corrado, and Jeff Dean.
\newblock Distributed representations of words and phrases and their
  compositionality.
\newblock In {\em Advances in Neural Information Processing Systems 26}, pages
  3111--3119. Curran Associates, Inc., 2013.

\bibitem{scikit}
Fabian Pedregosa, Ga\"{e}l Varoquaux, Alexandre Gramfort, Vincent Michel,
  Bertrand Thirion, Olivier Grisel, Mathieu Blondel, Peter Prettenhofer, Ron
  Weiss, Vincent Dubourg, Jake Vanderplas, Alexandre Passos, David Cournapeau,
  Matthieu Brucher, Matthieu Perrot, and \'{E}douard Duchesnay.
\newblock Scikit-learn: Machine learning in python.
\newblock {\em J. Mach. Learn. Res.}, 12:2825--2830, 2011.

\bibitem{Sakai:2014:SRI:2641383.2641385}
Tetsuya Sakai.
\newblock Statistical reform in information retrieval?
\newblock {\em SIGIR Forum}, 48(1):3--12, June 2014.

\bibitem{Schewe2014101}
Klaus-Dieter Schewe and Qing Wang.
\newblock A theoretical framework for knowledge-based entity resolution.
\newblock {\em Theoretical Computer Science}, 549:101 -- 126, 2014.

\bibitem{markov}
P.~Singla and P.~Domingos.
\newblock Entity resolution with markov logic.
\newblock In {\em Data Mining, 2006. ICDM '06. Sixth International Conference
  on}, pages 572--582, 2006.

\bibitem{yago}
Fabian~M. Suchanek, Gjergji Kasneci, and Gerhard Weikum.
\newblock Yago: A core of semantic knowledge.
\newblock In {\em Proceedings of the 16th International Conference on World
  Wide Web}, pages 697--706. ACM, 2007.

\bibitem{drmrer}
Huzaifa Syed, John Talburt, Fan Liu, Daniel Pullen, and Ningning Wu.
\newblock Developing and refining matching rules for entity resolution.
\newblock In {\em Proceedings of the International Conference on Information
  and Knowledge Engineering (IKE)}, pages 1--6, 2012.

\bibitem{book2}
J.R. Talburt.
\newblock {\em Entity Resolution and Information Quality}.
\newblock Elsevier Science, 2011.

\bibitem{Vechtomova201071}
Olga Vechtomova.
\newblock Facet-based opinion retrieval from blogs.
\newblock {\em Information Processing \& Management}, 46(1):71 -- 88, 2010.

\bibitem{crowder}
Jiannan Wang, Tim Kraska, Michael~J. Franklin, and Jianhua Feng.
\newblock Crowder: Crowdsourcing entity resolution.
\newblock {\em Proc. VLDB Endow.}, 5(11):1483--1494, 2012.

\bibitem{esjndd}
Chuan Xiao, Wei Wang, Xuemin Lin, and Jeffrey~Xu Yu.
\newblock Efficient similarity joins for near duplicate detection.
\newblock In {\em Proceedings of the 17th International Conference on World
  Wide Web}, pages 131--140. ACM, 2008.

\bibitem{Zhang:2015:TEC:2740908.2742766}
Lei Zhang, Yunpeng Dong, and Achim Rettinger.
\newblock Towards entity correctness, completeness and emergence for entity
  recognition.
\newblock In {\em Proceedings of the 24th International Conference on World
  Wide Web}, WWW '15 Companion, pages 143--144, New York, NY, USA, 2015. ACM.

\bibitem{Zhao2016}
Gang Zhao, Ji~Wu, Dingding Wang, and Tao Li.
\newblock Entity disambiguation to wikipedia using collective ranking.
\newblock {\em Information Processing \& Management}, 2016.
\newblock In Press.

\end{thebibliography}

\end{document}